\documentclass{sigchi-ext}

\usepackage[T1]{fontenc}
\usepackage{textcomp}
\usepackage[scaled=.92]{helvet} 
\usepackage{graphicx} 
\usepackage{balance}  
\usepackage{booktabs} 
\usepackage{ccicons}  
\usepackage{ragged2e} 

\def\plaintitle{Why Don't You Do Something About It? Outlining Connections between AI Explanations \& User Actions} 
\def\emptyauthor{}
\def\plainkeywords{explainable AI; actionability; human-centered design}

\title{Why Don't You \underline{Do} Something About It? Connecting AI Explanations \& User Action}

\numberofauthors{2}
\author{%
  \alignauthor{%
    \textbf{Gennie Mansi}\\
    \affaddr{Georgia Institute of Technology} \\
    \affaddr{Atlanta, GA}\\
    \email{gmansi3@gatech.edu}}\alignauthor{%
    \textbf{Mark Riedl}\\
    \affaddr{Georgia Institute of Technology}\\
    \affaddr{Atlanta, GA}\\
    \email{riedl@cc.gatech.edu}}}

\definecolor{linkColor}{RGB}{6,125,233}
\hypersetup{%
  pdftitle={\plaintitle},
  pdfauthor={\emptyauthor},
  pdfkeywords={\plainkeywords},
  bookmarksnumbered,
  pdfstartview={FitH},
  colorlinks,
  citecolor=black,
  filecolor=black,
  linkcolor=black,
  urlcolor=linkColor,
  breaklinks=true,
}

\begin{document}

\CopyrightYear{2023}
\setcopyright{rightsretained}
\conferenceinfo{CHI'23,}{April  25--30, 2023, Honolulu, HI, USA}
\isbn{978-1-4503-6819-3/20/04}
\doi{https://doi.org/10.1145/3334480.XXXXXXX}

\copyrightinfo{\acmcopyright}

\maketitle

\RaggedRight{}

\begin{abstract}
  A core assumption of explainable AI systems is that explanations change what users know, thereby enabling them to act within their complex socio-technical environments. Despite the centrality of action, explanations are often organized and evaluated based on technical aspects. Prior work varies widely in the connections it traces between information provided in explanations and resulting user actions. An important first step in centering action in evaluations is understanding what the XAI community collectively recognizes as the range of information that explanations can present and what actions are associated with them. In this paper, we present our framework, which maps prior work on information presented in explanations and user action, and we discuss the gaps we uncovered about the information presented to users. 
\end{abstract}

\keywords{\plainkeywords}

\ccsdesc[500]{Human-centered computing~User centered design}

\printccsdesc

\section{Introduction \& Background}

Artificial intelligence systems are increasingly involved in high-stakes decision making, such as healthcare, financial, and educational systems~\cite{Lipton2018, ChariEtAl2020, GuidottiEtAl2018, DohnEtAl2020}. Many have called for explainable AI (XAI)---AI systems that provide explanations for their reasoning or responses in a way that humans can understand~\cite{Lipton2018, ChariEtAl2020, GuidottiEtAl2018}. Through the explanations, developers and researchers aim to create systems that support user trust \cite{GuidottiEtAl2018, MothilalEtAl2020, FerrarioEtAl2020, Spiegelhalter2020, GliksonAndWoolley2020}, transparency \cite{GuidottiEtAl2018, GliksonAndWoolley2020, Krishnan2019, AndradaEtAl2022}, and control and agency \cite{AndradaEtAl2022, Slota2020, FriedmanAndKahn1992, DiblasiEtAl2020, Liu2021, CemalogluEtAl2019, FanniEtAl2022, Silva2019}. 

Explanation systems are often organized and evaluated based on technical aspects. For example, many papers \cite{AdadiAndBerrada2018, BiranAndCotton2017, CarvalhoEtAl2019, MohseniEtAl2020, ArrietaEtAl2019} organize and address explainability based on families of methods that are either inherently explainable (e.g. rule-based models, decision trees) or help with post-hoc explainability (e.g. Partial Dependency Plots, counterfactuals, sensitivity analysis). Some have noted the overall lack of papers evaluating XAI methods and quantifying their relevance \cite{SamekEtAl2019}, and those proposing models often focus on technical quantiative measures \cite{AdadiAndBerrada2018, ArrietaEtAl2019, MohseniEtAl2020}, such as accuracy when performing a task \cite{ArrietaEtAl2019, MohseniEtAl2020, SamekEtAl2019}, F1, and sensitivity \cite{ArrietaEtAl2019}.

While technical aspects are important for verifying algorithmic correctness, they don't speak to the core assumption of XAI systems: explanations change what users know, thereby enabling them to act in response to AI decisions~\cite{Liu2021, Silva2019, Coeckelbergh2020}. The terms \textbf{actionability} and \textbf{actionable insight} are used to reference explanations' abilities to enable pragmatic action via the information they provide \cite{JornoAndGynther2018, LyuEtAl2016, ChoEtAl2019, JoshiEtAl2019, TanAndChan2016, WiratungaEtAl2021, SinghEtAl2021}. However, Liao, Gruen, and Miller~\cite{LiaoEtAl2020} note the gap in delivering satisfying user experiences. Researchers are increasingly calling for greater consideration of users' actions when designing and evaluating XAI systems~\cite{JornoAndGynther2018,FanniEtAl2022}. Similarly, Ehsan et al. \cite{EhsanEtAl2021} propose incorporating social transparency---making visible socio-organizational factors that govern AI to help users more effectively take action.

Not only is there no consensus on what kinds of information can and should be presented to users, but there is also a relative lack of connection between types of explanations and what the resulting user actions are expected to be. Several papers~\cite{Lipton2018, GuidottiEtAl2018, Doshi-VelezAndKim2017, WangEtAl2019, AryaEtAl2019, ICOAndTuring2022, ChariEtAl2020} attempt to categorize information that explanations communicate, but many don't elaborate expected user actions. Others make a connection with user action by evaluating explanations on users' ability to act on unexpected behaviors \cite{ChoEtAl2019}, change an outcome \cite{MothilalEtAl2020, JoshiEtAl2019, ChiangAndDey2018}, or achieve a desired result \cite{SinghEtAl2021, EnsanEtAl2017, GhoshEtAl2018}. But these evaluate specific technical solutions.  

We argue that the evaluation of the effectiveness of explanations be tied directly to the actions a user can take in response. Establishing action as a new heuristic for judging explanations is key to forcing designers to consider the explanation's integration with users' broader system factors. However, the lack of consensus on information that XAI systems present and what corresponding user actions one should want or expect can make the design and evaluation of XAI systems challenging.

We help reorient the evaluation of explanations by creating a framework that attempts to tie information presented by XAI systems to user action. We use existing literature to synthesize 10 categories of information that can be presented by XAI systems. For each category, we explicitly list actions users can take in response. Through our framework, we provide a more unified starting place to examine the relationship between information provided in explanations, user actions, and ultimate XAI interaction goals of trust, transparency, control, and agency.

\section{Creating the Framework}

Using existing XAI literature, we consolidated a framework mapping the kinds of information in explanations to associated actions. Papers focused on cataloging technical methodologies as opposed to information in explanations were excluded. After reading the abstracts in the search results, the authors selected 30 survey papers for deeper reading. Of these, 11 met the inclusion criteria: \cite{ICOAndTuring2022, ChariEtAl2020, GuidottiEtAl2018, LimEtAl2019, Guidotti2020, Lipton2018, EdwardsAndVeale2017, LiaoEtAl2020, ArrietaEtAl2019, Yao2021, GilpinEtAl2018}. We iteratively developed the categories in our framework. We began with the categories listed out by the ICO \& Turing Institute \cite{ICOAndTuring2022}, cataloguing information types and any actions mentioned as intended or potentially resulting from the information provided. If a paper included a type of information that could not be clearly classified, we redefined our framework's categories, and then re-classified all information types and corresponding actions. Our framework isn't exhaustive, but a starting point.

\section{The Framework}
\begin{figure}
\vspace*{-3.5cm}\hspace{-5cm}
\begin{minipage}{3\columnwidth}
       \includegraphics[width=\linewidth]{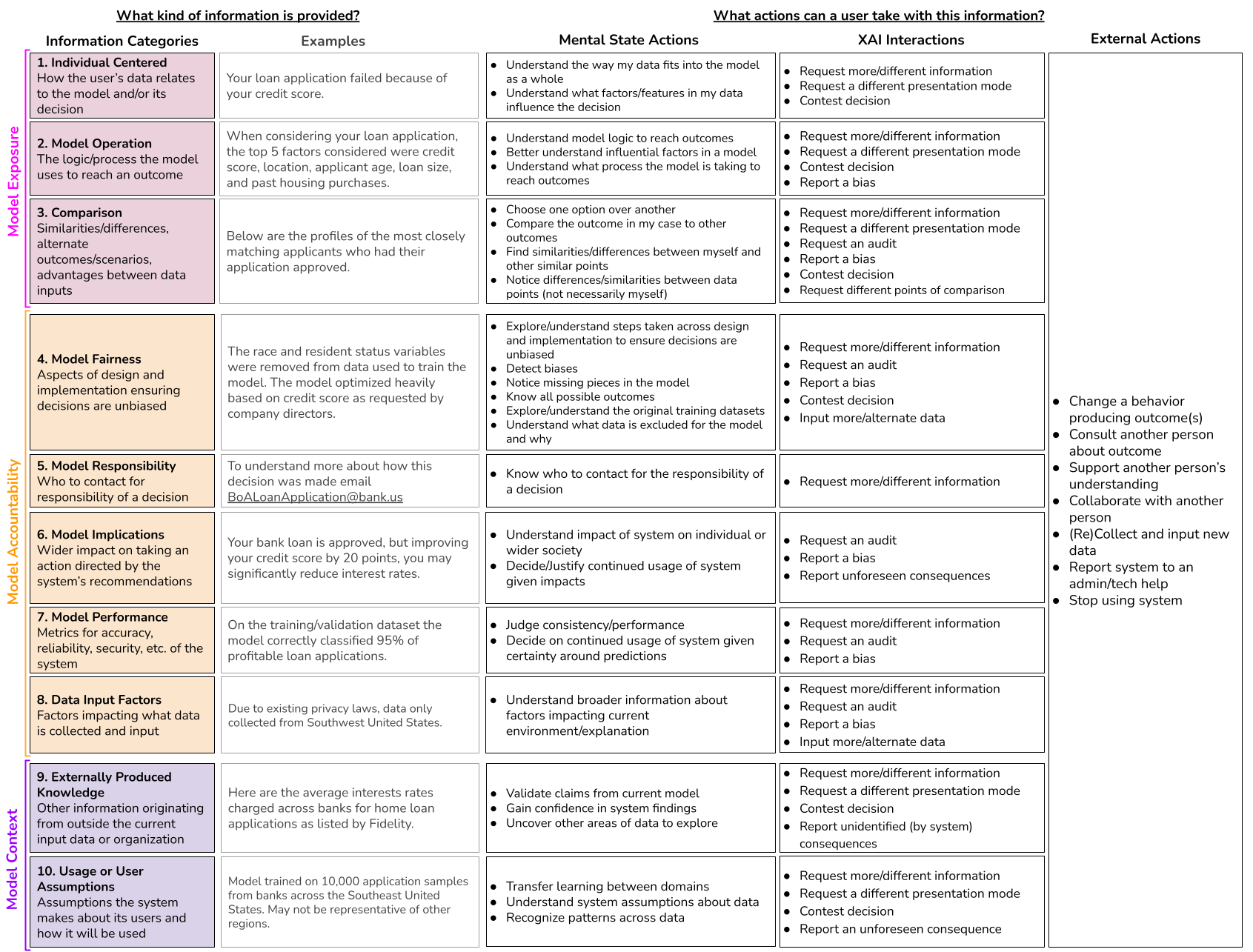} 
       \caption{The Framework maps 10 categories of information XAI systems contain to actions distinguished by how users interact with their environment.} 
       \label{fig:framework} 
\end{minipage} 
\end{figure}

Our Framework, displayed in Figure \ref{fig:framework}, has 10 categories of information that are thematically grouped. Explanations with information on \textbf{Model Exposure} (pink) categories communicate how the model relates to the AI's decision to the user. Explanations with providing information on \textbf{Model Accountability} (orange) communicate about the creation, verification, and implications of the AI model. 
Finally, explanations providing information on \textbf{Model Context} (purple) communicate how the model aligns with externally produced information and its assumptions about users' context. 

To the right of each category we list actions users can take in response. Jørnø \& Gynther  \cite{JornoAndGynther2018} and Fanni et al. \cite{FanniEtAl2022} argue that an actor's ``action capabilities'' encompass mental actions or choices, which can impact physical actions or interactions. 
Thus, actionability can manifest as mental or physical actions. Consequently, we define three classes of actions. \textbf{(1)~Mental State Actions} are changes in users' mental state that impact what they understand or believe about the system and how they can act in response. \textbf{(2)~XAI Interactions} are the interactions users have with the XAI interface features, such as clicking buttons or toggles to request more information. Finally, \textbf{(3)~External Actions} are actions users can take in response to and outside of the XAI system, such as reaching out to another person for help or ceasing to use the system. We acknowledge Mental State Actions are not traditionally considered part of ``actionability``. However, they constitute responses that potentially influence XAI Interactions and External Actions. For example, deciding on the continued usage of a system is neither an XAI Interaction nor an External Action but can impact both. Consequently, Mental State Actions are an important part of actionability that need to be considered. 

While the user has a broad range of explanation types and resulting Mental State Actions, there are surprisingly limited potential XAI Interactios or External Actions. There are only 7 distinct XAI Interactions and 6 distinct External Actions. Further, the distribution of papers presenting explanations with information types across the typology revealed several gaps. Almost every paper (12 total) mentioned explanation(s) with Individual-Centered, Modus Operandi, Comparison, and Model Fairness information. However, only about half (4-6) of papers mentioned explanations with Model Performance and Model Input Environment information. Explanation examples with information on Model Implications, Model Responsibility, Externally Produced Knowledge, and Usage Assumptions were mentioned in no more than 2 papers. This may indicate a need for AI creators to expand the kinds of information and actions they can provide to users. 

\section{Conclusions}

A core underlying assumption is that explanations are useful to users and that usefulness consequently manifests itself in the potential for action, both mental and physical. Through the examination and synthesis of existing literature, our framework contributes a first step in developing transferable evaluations centered on assessing explanations based on users' actions, and we uncovered several gaps in the literature around the kinds of information provided by explanations. Addressing the gaps may result in a richer design space of XAI systems that afford a greater range of actionability to users. The framework places actionability as the locus of successful XAI systems and suggests new opportunities to evaluate the effectiveness of explanations as an increase in the action potential of users.

\clearpage

\clearpage

\balance{} 

\section{Acknowledgements}
This material is based upon work supported by the National Science Foundation GRFP under Grant No. DGE-2039655. Any opinion, findings, and conclusions or recommendations expressed in this material are those of the authors(s) and do not necessarily reflect the views of the National Science Foundation.
Special thanks to Kaely Hall.

\bibliographystyle{SIGCHI-Reference-Format}
\bibliography{main}

\end{document}